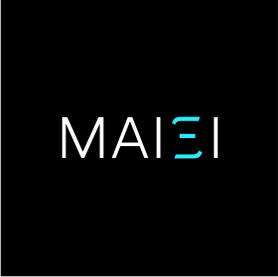

**Montreal AI Ethics Institute**

*Democratizing AI Ethics Literacy*

https://montrealethics.ai

# THE CO-DESIGNED POST-PANDEMIC UNIVERSITY: A PARTICIPATORY AND CONTINUAL LEARNING APPROACH FOR THE FUTURE OF WORK

A conference submission for the Post-Pandemic University prepared by Abhishek Gupta and Connor Wright

Published on September 6th, 2020

*Primary contact for the report:*

Connor Wright (Connor@montrealethics.ai)

Partnerships Manager, Montreal AI Ethics Institute

The pandemic has shattered the traditional enclosures of learning. The post-pandemic university (PPU) will no longer be contained within the 4 walls of a lecture theatre, and finish once students have left the premises. The use of online services has now blended home and university life, and the PPU needs to reflect this. Our proposal of a continuous learning model will take advantage of the newfound omnipresence of learning, while being dynamic enough to continually adapt to the ever-evolving virus situation. Universities restricting themselves to fixed subject themes that are then forgotten once completed, will miss out on the 'fresh start' presented by the virus.

This launch pad gives us the chance to propose the accompaniment of participatory learning. Here, the reliance on online services increases learning's reach to students via the geographical barrier being removed, helping them to participate more in the knowledge being cultivated. No longer will knowledge be presented to the students in a pre-formed structure by the lecturer, but rather the students are supplemented by said structure, subsequently adding their own 'nuts and bolts'. Resultantly, the formation of the constant learning output is guided by the students' insights through our proposed framework of participatory learning.

As the virus has shown, our learning can be affected in ways never before seen. Our proposal of continual learning adapts to the unrestricted learning space, while taking advantage of the increased participation afforded by the subsequent emphasis on online learning that will be the hallmarks of the PPU.

**Why participatory learning?**

Participatory learning, as characterized by Domínguez (Domínguez, 2012), places students at the center of the learning process. Now, students will have more say in the curriculum, allowing for adaptability to topics that matter to the students and educators, and course adaptation to the constantly changing pandemic arena. With topics that interest the students now brought to the fore, so is an increased level of engagement. The curriculum creates a sense of ownership by individual students that is to be cared for, nurtured and enjoyed thanks to the involvement of topics dear to the students. As a result, this provides a unique learning opportunity to learn from different peers. There is no better topic to explain than one you are passionate about, and no better teacher of a topic than someone who cares. Our work on participatory learning thus simultaneously provides unique perspectives, while increasing student participation in the PPU.

This increased engagement presents another unique opportunity for the PPU through participatory learning, namely focusing on the application of knowledge. Co-authoring the curriculum means students can bring in material that pertains to the effects the pandemic is having over different subject areas, both novel and traditional. Mathematical lectures can now feature sessions on algorithms being used in contact-tracing apps, while the issues of privacy contained within this topic can be tackled by philosophers. Students' views on such apps and how paramount such discussions are in today's environment would be lost within the old education system, and the PPU has the chance to rectify this through participatory learning.

**Why continual learning?**

Many arguments have been put forward on the uncertainty regarding the future of work (Howard, 2019), especially as it relates to what skills should be taught in the educational system (Herold, 2017) and reskilling programs (Illanes, et al., 2018) offered in continuing education and corporate ecosystems. One thing that is certain, something that will help new entrants into the workforce and existing workers not only survive, but thrive in this new landscape is the ability to keep learning. The pandemic is helping to accelerate that trend and snap into focus the requirement that the university and educational model be transformed from mapping static, predetermined educational requirements (Arlett (Assistant Director), et al., 2010) to dynamic, evolving, and co-developed curricula that constantly adapt to meet the changing requirements of industry. In our work, we see continual learning as both the supply of constantly evolving educational materials (which admittedly will require substantial work on the part of educators, perhaps calling for the need to separate researchers from lecturers a lot more in the existing university ecosystems) and more importantly the recognition that a shift in mindset is required on the part of learners. This shift in mindset is one that eschews complacency in favor of an infant-like curiosity applied to their craft, where the word craft is used deliberately to evoke a sense of continual tinkering and adaptation that is not achieved in the current one-way adsorption (a surface-level absorption learning) of learning material. A true craftsman is someone who identifies that thriving in the new post-pandemic world, where they are placed in competition not just in the local ecosystem but with everyone because of the possibilities of remote hiring and work, requires them to hone their craft at the altar of continual learning.

**How does continual and participatory learning work together?**

Having presented our two accounts, now is the time to harmonize. Continual learning allows students at the PPU to keep up to date with the ever-changing pandemic scenario, while participatory learning surfaces the new trends within. Here, partaking in current topics allows for the previously mentioned continual honing of the skills becoming desirable by different sectors, and by the pandemic environment. Thus, relevancy is key, and this will be achieved through adopting an iterative approach.

This approach can be best explained through an analogy of a machine. The overall machine can be seen as the entire course (such as Philosophy), made up of large individual parts (differing modules). Previously, modules were perceived as large parts of the machine, taking lots of consideration and research to replace. However, an iterative approach rather treats the machine as made of much smaller individual parts which we dub micromodules. Accordingly, parts are more easily replaceable without upsetting the whole course ecosystem and require less workload to replace. Relevancy can then be maintained to the highest standard through modules being easy to swap in and out, while fomenting an agile mindset amongst students.

To test this approach, we propose assessments aligned with industry standards. Paired with the idea of relevancy, examinations will need to adapt to the agile content while also adjusting to the new industry environment created by the pandemic. Examinations being aligned with industry allows for students to apply their knowledge learnt to the situation at hand, while also adhering to the new insights provided on the constantly updated content.

**An experiment at the Montreal AI Ethics Institute via the learning communities**

The learning communities at MAIEI are driven by this very ideology. They are focused on 5 areas: complex systems theory, privacy, disinformation, machine learning security, and labor impacts of AI. The syllabi are seeded with starting points but are iterated upon by the participants of the community. It is structured in a flipped classroom model that encourages collaborative discussions over the pedantic explanations of the source material. The participants are encouraged to explore the material first and then build on that learning with others in the community. Given that there is a loose structure provided to the participants, it encourages serendipitous connections between other material, inviting participants to bring in their lived experiences and backgrounds to the discussion which is something that is not possible in a traditional classroom. The remote delivery mechanism also allows people from diverse geographies to come together in ways that would not otherwise be possible, given financial, time, and other barriers to participation.

**What does the future hold with these approaches?**

The future is constantly being redefined by the effects of the pandemic, and it is the purpose of the PPU to maintain students' part in it. Aligning course material via participatory learning will allow the PPU to keep up-to-date with what interests the students, and continual learning will serve to maintain that this interest is in line with evolving industry trends. To further ensure this, the iterative approach will foster the rapid acquisition of new skills through its swapping in and out of modules, keeping in line with what skills are being desired by industry professionals. The engagement raised by participatory learning ensures that this adaptive approach will not fall on deaf ears, while the surfacing of new and unique insights proves the welcomed stepping stones in navigating the constantly changing pandemic-fueled environment.

We take an optimistic view on what the future holds, specifically we think it is essential that educators work hand-in-hand with learners who, increasingly so with the recent push in the pandemic, are able to leverage their networks and reach out to target audiences who work in the industries that they see themselves gravitating towards. This gives them real-time, on-the-ground intelligence in terms of what is needed to make themselves more adept to fit industry. Educators can certainly benefit from this intelligence by inviting these learners into the curriculum design process. Secondly, through such a collaborative approach, they are able to spread the benefits of this intelligence to other students as well who may not be as resourceful in finding this knowledge on their own.

Using micro-modules as the go-to methodology for teaching will empower educators to quickly adapt content and teaching strategies in response to the rapidly changing environment. They will also have the ability to garner feedback from learners, which is incentivized through the shorter format which is less burdensome, that will further empower them to adapt content and teaching methodologies that better equip learners for industry and whatever else they choose to embark on after their educational journey.

Finally, we believe this harmonized participatory and continual learning approach creates the perfect mix that will equip learners for the future of work with the current pandemic induced environment providing a timely sandbox to test-run this approach in an applied context.